\documentstyle[12pt]{article}
\begin{document}

\author{M. I. Krivoruchenko \\
%EndAName
{\small {\it Institute for Theoretical and Experimental Physics,
B.Cheremushkinskaya 25}}\\
{\small {\it 117259 Moscow, Russia}}}
\title{Explicit Solutions to the Unitarity Relations for Isovector Nucleon Form
Factors}
\date{}
\maketitle

\begin{abstract}
The explicit one-dimensional integral representations for isovector nucleon
form factors are constructed by solving the unitarity relations in terms of
the pion form factor and imaginary parts of the $t$-channel $p$-wave $\pi N$%
-scattering amplitudes. The possibility of computation of the high-energy
parts of the integrals with the use of the information on the low-energy
real parts of the $\pi N$-scattering amplitudes is discussed.
\end{abstract}

\vspace{1 cm}

\section{Introduction}

\setcounter{equation}{0}

The unitarity relations for isovector nucleon form factors have been
discovered by Frazer and Fulco$^1$ in 1960 yr. These relations were
discussed intensively in the 60-th and 70-th years. At present, new precise
data are available at low and intermediate momentum transfers both at
spacelike and timelike regions (see {\it e.g.} Ref.2 and references
therein). It is desirable therefore to analyze more carefully theoretical
methods and current models. In Sect.2, we report new analytical results
concerning the unitarity relations. In Sect.3, these results are used to
develop an alternative approach for numerical evaluation of the nucleon form
factors from the unitarity relations.

\section{Solutions to the Unitarity Relations}

\setcounter{equation}{0}

The unitarity relations for isovector nucleon form factors have the form$^1$

\begin{equation}
\label{I}ImF_{iv}(t)=-\frac{k^3}{\sqrt{t}}F_\pi ^{*}(t)\Gamma _i(t) 
\end{equation}
where $k=\sqrt{t/4-\mu ^2}$, $\mu $ is the pion mass, $F_\pi (t)$ is the
pion form factor, and $\Gamma _i(t)$ are combinations of the $t$-channel $p$%
-wave projections of the $\pi N$-scattering amplitudes. The form factors are
normalized by $F_{1v}(0)=1/2$ and $F_{2v}(0)=(\mu _p-\mu _n-1)/2$ with $\mu
_p$ and $\mu _n$ being the proton and neutron magnetic moments (in n.m.).
The proton charge is set equal to unity.

In the approximation of neglecting all but the two-pion intermediate state,
the values $\Gamma _i(t)$ obey the once subtracted dispersion relations 
\begin{equation}
\label{II}\Gamma _i(t)/F_\pi (t)=Re\frac{\Gamma _i(\kappa )}{F_\pi (\kappa )}%
+\frac{t-\kappa }\pi P\int_{-\infty }^a\frac{Im\Gamma _i(t^{\prime })}{%
(t^{\prime }-\kappa )(t^{\prime }-t)}\frac{dt^{\prime }}{F_\pi (t^{\prime })}%
. 
\end{equation}
with $a=4\mu ^2-\mu ^4/m^2$. In the FFGS model$^{1,3}$, the following
relation takes place above the two-pion threshold 
\begin{equation}
\label{III}\frac{k^3}{\sqrt{t}}|F_\pi (t)|^2=D(0)\frac{k_\rho ^3}{m_\rho
^2\Gamma _\rho }\frac{t_0+t}{t_0}ImF_\pi (t+i0). 
\end{equation}
where $D(0)$ is the value of the $D$-function at the origin. The
exponentially large parameter $t_0$ is a unique root of equation $D(-t_0)=0$%
, $k_\rho $ is the pion momentum at $t=m_\rho ^2$, $m_\rho $ and $\Gamma
_\rho $ are the $\rho $-meson mass and width. In the narrow width limit, $%
D(0)=m_\rho ^2$, $t_0=\infty $ .

Combining Eqs.(2.1), (2.2) and (2.3), the once subtracted dispersion
integral for the nucleon form factors can be evaluated to give

$$
F_{iv}(t)=F_{iv}(-t_0)-D(0)\frac{k_\rho ^3}{m_\rho ^2\Gamma _\rho }\frac{%
t_0+t}{t_0}F_\pi (t) 
$$
\begin{equation}
\label{IV}(Re\frac{\Gamma _i(\kappa )}{F_\pi (\kappa )}+\frac 1\pi
P\int_{-\infty }^a\frac{Im\Gamma _i(t^{\prime })}{(t^{\prime }-\kappa )}%
\frac{(t-\kappa )/F_\pi (t^{\prime })-(t^{\prime }-\kappa )/F_\pi (t)}{%
(t^{\prime }-t)}dt^{\prime }). 
\end{equation}

For isovector kaon form factor, similar representation has been constructed
within the differential approximation scheme$^4$. For computation of the
nucleon form factors, it is sufficient to know the imaginary parts $Im\Gamma
_i(t)$ at the interval $(-\infty ,a)$ only.

The imaginary part of the ratio $Im{F}_{{iv}}{(t)/F(t)}$ has a simple form 
\begin{equation}
\label{V}Im\{(F_{iv}(t)-F_{iv}(-t_0))/F_\pi (t)\}/\frac{k^3}{\sqrt{t}}%
=-\frac 1\pi \int_{-\infty }^a\frac{Im\Gamma _i(t^{\prime })}{t^{\prime }-t}%
dt^{\prime }. 
\end{equation}

\section{High-Energy Part of the Integrals}

\setcounter{equation}{0}

The integrals in Eqs.(2.4) and (2.5) are determined by the low- and
high-energy parts of the values $Im\Gamma _i(t)$. The values $\Gamma _i(t)$
can be decomposed to the Born and rescattering parts: $\Gamma _i(t)=\Gamma
_{iB}(t)+\Gamma _{iR}(t)$. The values $\Gamma _{iR}(t)$ are known at the
interval $-26\mu ^2<t<0$ from the phase-shift analysis of the $\pi N$%
-scattering amplitudes. The high-energy rescattering parts $\Gamma _{iR}(t)$
at $-\infty <t<-26\mu ^2 = b$ are unknown. The problem 
consists in evaluation of
the high-energy parts with the help of the information on the low-energy
real pars $Re\Gamma _{iR}(t)$.

In the ratio $Re{F}_{{iv}}{(t)/F}_\pi {(t)}$, the convergence of the
integral can be improved by taking a weighted sum%
$$
Re\{(F_{iv}(t)-F_{iv}(-t_0))/F_\pi (t)\}=-D(0)\frac{k_\rho ^3}{m_\rho
^2\Gamma _\rho }\frac{t_0+t}{t_0} 
$$
\begin{equation}
\label{VI}\sum_{j=1}^Nc_{ij}(Re\frac{\Gamma _i(\kappa _j)}{F_\pi (\kappa _j)}%
+\frac 1\pi P\int_{-\infty }^a\frac{Im\Gamma _i(t^{\prime })}{(t^{\prime
}-\kappa _j)}\frac{(t-\kappa _j)/F_\pi (t^{\prime })-(t^{\prime }-\kappa
_j)Re\{1/F_\pi (t)\}}{(t^{\prime }-t)}dt^{\prime }). 
\end{equation}
with the coefficients $\sum_{j=1}^Nc_{ij}=1$. The subtraction points $\kappa
_j$ are assumed to belong to the interval $-26\mu ^2<\kappa _j<0$. The
coefficients $c_{ij}$ must be chosen such as to minimize contributions to
the integral from the high-energy part $-\infty <t<-26\mu ^2$.

The convergence of the integral in Eq.(2.5) can be improved in the similar
way. The right hand side of Eq.(2.5) can be rewritten as follows 
\begin{equation}
\label{VII}\sum_{j=1}^Nd_{ij}Re\frac{\Gamma _i(\kappa _j)}{F_\pi (\kappa _j)}%
-\frac 1\pi (\int_{-\infty }^b+P\int_b^a)Im\Gamma _i(t^{\prime
})(\sum_{j=1}^Nd_{ij}\frac 1{t^{\prime }-\kappa _j}\frac 1{F_\pi (t^{\prime
})}+\frac 1{t^{\prime }-t})dt^{\prime } 
\end{equation}
The weight coefficients $d_{ij}$ must be chosen such as to suppress the
high-energy part of the integral. To improve convergence of the integral,
one should require $\sum_{j=1}^Nd_{ij}=0.$

\section{Conclusion}

In the standard approach$^1$, one should compute two integrals numerically.
The first one comes from the dispersion integral for the $\Gamma _i(t)$. The
second one comes from the dispersion integral for the nucleon form factors.
We showed that the last integral can be evaluated analytically. In doing so,
we arrived to the explicit one-dimensional integral representations (2.4)
for the nucleon form factors in terms of the pion form factor and the
amplitudes $\Gamma _i(t)$ at $t<a$. These representations provide an
alternative possibility for numerical analysis of the problem.

\end{document}